\documentclass[twocolumn,preprintnumbers,superscriptaddress,nofootinbib]{revtex4}
\usepackage{graphicx}
\usepackage{amssymb}
\usepackage{color}
\usepackage{enumerate}
\def\beq{\begin{equation}}
\def\eeq{\end{equation}}
\def\beqn{\begin{eqnarray}}
\def\eeqn{\end{eqnarray}}

\renewcommand{\texttt}{{}}
\newcommand{\be}{\begin{eqnarray}}
\newcommand{\ee}{\end{eqnarray}}
\newcommand{\bee}{\begin{equation}}
\newcommand{\eee}{\end{equation}}

\begin{document}

\title{Super-renormalizable or finite completion
of the Starobinsky theory}

\author{Fabio Briscese}
\email{fabio.briscese@sbai.uniroma1.it}
\affiliation{Istituto Nazionale di Alta Matematica Francesco
Severi, Gruppo Nazionale di Fisica Matematica,
Citt\`a Universitaria, P.le A. Moro 5, 00185 Rome, EU}
\affiliation{Dipartimento SBAI, Sezione di Matematica, Sapienza Universit\`a di Roma,
Via Antonio Scarpa 16,  00161 Rome, EU}

\author{Leonardo Modesto}
\email{lmodesto@fudan.edu.cn}
%\email{lmodesto@perimeterinstitute.ca}
\affiliation{{\small Department of Physics \& Center for Field Theory and Particle Physics,} \\
{\small Fudan University, 200433 Shanghai, China}
}

\author{Shinji Tsujikawa}
\email{shinji@rs.kagu.tus.ac.jp}
\affiliation{Department of Physics, Faculty of Science, Tokyo University of Science,
1-3, Kagurazaka, Shinjuku-ku, Tokyo 162-8601, Japan}

\date{\small\today}

\begin{abstract} \noindent

The recent Planck data of Cosmic Microwave Background (CMB)
temperature anisotropies support the Starobinsky theory
in which the quadratic Ricci scalar drives cosmic inflation.
We build up a multi-dimensional quantum consisted ultraviolet
completion of the model in a phenomenological ``bottom-up approach".
We present the maximal class of theories compatible
with unitarity and (super-)renormalizability or finiteness which reduces to
the Starobinsky theory in the low-energy limit.
The outcome is a maximal extension of the Krasnikov-Tomboulis-Modesto theory
including an extra scalar degree of freedom besides the graviton field.
The original theory was afterwards independently discovered by Biswas-Gerwick-Koivisto-Mazumdar
starting from first principles.
We explicitly show power counting super-renormalizability or
finiteness (in odd dimensions) and unitarity (no ghosts) of the theory.
Any further extension of the theory is non-unitary confirming the existence
of at most one single extra degree of freedom, the scalaron.
A mechanism to achieve the Starobinsky theory in string (field) theory
is also investigated at the end of the paper.

\end{abstract}

\maketitle

\section{Introduction}

Inflation not only resolves a number of cosmological puzzles plagued
in the standard big bang cosmology \cite{Starobinsky,oldinf},
but also it can account for the origin
of large-scale structure in the Universe \cite{oldper}.
The standard slow-roll single-field inflationary scenario gives rise to
nearly scale-invariant and adiabatic primordial perturbations by stretching
out quantum fluctuations over super-Hubble scales.
This prediction was confirmed by the COBE \cite{COBE} and
WMAP \cite{WMAP1} groups from the observations
of the CMB temperature anisotropies.

Recently, the Planck group provided high-precision data
of the CMB temperature anisotropies \cite{Planck1,Planck2,Planck3,Planck4},
by which the spectral index $n_s$ of curvature perturbations, the tensor-to-scalar
ratio $r$, the non-linear parameter $f_{\rm NL}$ of primordial
non-Gaussianities were tightly constrained relative to the bounds
derived by the WMAP 9-year data \cite{WMAP9}.
Since these observables are different depending on the models of
inflation \cite{review}, we can discriminate between
a host of inflationary models from the Planck data.
In particular the bound of the non-linear estimator of the squeezed shape
is $f_{\rm NL}^{\rm local}=2.7 \pm 5.8$ (68~\%\,CL) \cite{Planck4},
by which all of the single-field slow-variation inflationary models are
consistent with the bound of local non-Gaussianities.

The joint data analysis combined with Planck and the WMAP polarization (WP)
measurement shows that the scalar spectral index is constrained to be
$n_s = 0.9603 \pm 0.0073$, by which the Harrison-Zel'dovich spectrum
is excluded at more than $5 \sigma$ CL \cite{Planck3}.
The tensor-to-scalar ratio is bounded to be $r<0.12$ (95~\%\,CL),
which corresponds to an upper bound for the inflationary energy scale
of about $(1.9 \times 10^{16}\,{\rm GeV})^4$.
In Ref.~\cite{Tsujikawa} the authors discriminated between single-field inflationary
models which belong to the class of most general scalar-tensor theories
with second-order equations of motion (Horndeski's theory \cite{Horndeski}).

There are many slow-roll single-field models which are tension with
the Planck data \cite{Planck3,Tsujikawa}.
For example, power-law inflation with the exponential potential
$V= \Lambda^4 \exp(-\lambda\phi/M_{\rm pl})$ \cite{powerlaw}
(where $M_{\rm Pl}=2.435 \times 10^{18}$\,GeV is the reduced Planck mass)
and chaotic inflation with the potential $V = \lambda \phi^n/n$ ($n>2$) \cite{chaotic}
are outside the $95\%$ CL boundary because of the large tensor-to-scalar ratio.
Even chaotic inflation with the powers $n=1,2/3$ \cite{mono} is
under an observational pressure due to the large scalar spectral index.
Hybrid inflation with the potential
$V=\Lambda^4+m^2 \phi^2/2+ \ldots$ \cite{hybrid}
gives a blue-tilted scalar spectrum ($n_s>1$) and hence
it is disfavored from the data.
Other models such as hill-top inflation with
$V = \Lambda ^4 ( 1 - \phi^2/\mu^2 + \ldots)$ \cite{newinf} or
natural inflation with $V = \Lambda^4 [1 + \cos(\phi/f)]$ \cite{natural}
are viable for a restricted range of parameters (e.g. $f \gtrsim 5
M_{\rm Pl}$ for natural inflation).

The models favored from the Planck data are
the ``small-field'' scenario in which the tensor-to-scalar ratio is
suppressed because of the small variation of the field during inflation
with the scalar spectral index $n_s \simeq 1-2/N$, where
$N=50$-60 is the number of e-foldings from the end
of inflation \cite{Tsujikawa}.
This includes the Starobinsky model with the Lagrangian
$f(R)=R+\epsilon R^2$ ($R$ is the Ricci scalar) \cite{Starobinsky}
and the non-minimally coupled model ($\xi R \phi^2/2$)
with the self-coupling potential $V(\phi)=\lambda \phi^4/4$ \cite{Futamase}
or the Higgs potential $V(\phi)=\lambda (\phi^2-\mu^2)^2/4$ \cite{Bezrukov}.
In fact, the resulting power spectra of scalar and tensor perturbations
in the Starobinsky's model is the same as those in the non-minimally
coupled model in the limit $|\xi| \gg 1$ \cite{fRper}.

In the Starobinsky model the quadratic curvature term $R^2$
drives inflation, which is followed by the gravitational reheating
with the decrease of $R^2$ \cite{Starreheating,Vilenkin,Suen,TMT}
(see Refs.~\cite{Sotiriou,Antonio,Clifton} for reviews).
If we make a conformal transformation to the Einstein frame, there
appears a scalar degree of freedom $\phi$
called ``scalaron'' \cite{Starobinsky} with a nearly flat potential
in the regime
$\phi \gg M_{\rm pl}$. The presence of the scalaron is crucial
to generate density perturbations consistent with the Planck data.
Recently there have been numerous attempts to construct
the Starobnisky model in the context of
supergravity \cite{Cecotti1987,Cecotti1988,Ketov,Watanabe,Ellis,Kallosh,sugra1,sugra2,sugra3,Ferrara2013}.

In this paper we go in search of an ultraviolet completion of
the Starobinsky's $f(R)$ action,
assuming a pure gravitational origin of inflation.
We start restricting our attention on the most general local
quadratic action for gravity \cite{Stelle,Shapiro,Shapirobook},
\bee
\mathcal{L}_{\rm quadratic} = R + \epsilon R^2 + \zeta \,
C_{\mu \nu \rho \sigma} C^{\mu \nu \rho \sigma}
+ \eta \,  \chi_E\,,
\eee
where $C_{\mu\nu\rho \sigma}$ is the Weyl tensor and
$\chi_E$ is the density of the Euler number.
Assuming a Friedmann-Lema\^{i}tre-Robertson-Walker
(homogeneous and isotropic) metric the Weyl term
vanishes because the metric is conformally flat and the
Lagrangian simplifies to the Starobinsky theory,
\bee
\mathcal{L}_{\rm Starobinsky} = R + \epsilon R^2. \label{Staro}
\eee

The goal of this paper is to find a (i) Lorentz invariant, (ii)
(super-)renormalizable or finite, (iii) unitary theory of gravity
which reduces to the Starobinsky theory in the low-energy limit. In
trying to achieve this goal we are looking for a {\em new
classical theory of gravity} which is {\em renormalizable at
quantum level}. This point of view opposes to
the emergent gravity scenery.

We can resume as follows the theoretical and observative
consistency requirements for a full quantum gravity theory.
\begin{enumerate}
\renewcommand{\theenumi}{\arabic{enumi}}
\item Unitarity (theoretical). A general theory is well defined if
``tachyons" and ``ghosts" are absent, in which case the
corresponding propagator has only first poles at $k^2 - M_i^2 =0$
with real masses (no tachyons) and with positive
residues (no ghosts).
\item Super-renormalizability or finiteness (theoretical). This
hypothesis makes consistent the theory at quantum level in analogy
with all the other fundamental interactions.
\item Lorentz invariance (observative). This is a symmetry of
nature well tested below the Planck scale.
\item The theory must be at least quadratic in the curvature
(observative) to achieve the agreement with the recent Planck data
confirming the viability of the Starobinsky theory.
\item The energy conditions are not violated from the matter side
(observative), but can be violated because higher-derivative
operators are present in the classical theory.
\end{enumerate}

Possible candidate theories which satisfy the above requirements
are listed below.
\begin{enumerate}
\renewcommand{\theenumi}{(\Roman{enumi})}
\renewcommand{\labelenumi}{\theenumi}
%(i)
\item We have several candidate theories, but only one
fulfills all the above requirements at perturbative level.
We call it: ``super-renormalizable or finite gravity" \cite{Krasnikov, Tomboulis, Modesto, M2, M3, M4,
Biswas}. This is the theory we are going to mainly concentrate on in this paper. Another candidate perturbatively renormalizable and unitary theory has been studied
in Ref.~\cite{Narain}.
\item
At non-perturbative level a natural candidate theory is ``asymptotic safe quantum gravity"
in which the mass of the ghost diverges when the momentum scale goes to infinity
and the mode decouples from the theory in the ultraviolet regime or the ghost pole
is moved further by the renormalization group running
\cite{Reuter, Percacci, Dario}.
\item String theory or its field redefinition could do the job because the modification
of the propagator coming from ``string field theory" \cite{SFT}
 makes possible to get the Starobinsky model in the low-energy limit.
We can easily extend  supergravity in ten dimensions incorporating the modification
suggested by string field theory together with diffeomorphism
invariance \cite{Calcagni-Modesto, Calcagni-Modesto-Nicolini, deser}.
We will come back on this point at the end of the paper.

The problem lies in the spectrum of the theory which contains extra
massless degrees of freedom besides the graviton field and an infinite tower of massive states.
It is not clear how to select out only one scalar degree of freedom to sustain inflation
and reproduce the correct perturbation spectrum,
while the other massive scalars will be suppressed
by a higher mass scale involving the volume of the compact space
or the extra dimensions in a brane-cosmology scenery.
However, a priori we cannot exclude this possibility
and we strongly suggest to investigate and engineer in this direction.

\end{enumerate}
%}

The last very important feature of the theory we are looking
for is the presence of one extra scalar degree of
freedom in the gravitational spectrum, without voiding unitarity
and renormalizability. We call this degree of freedom
gravi-scalar or scalaron, which is of
fundamental importance to generate primordial perturbations
during inflation.

A previous work \cite{M4} presented a super-renormalizable
and ghost-free theory of gravity, which, under
a natural exponential ansatz of the form factor and a suitable
truncation, gives rise to the Starobinsky model.
However the problem with such a model is
that, at quantum level, the linearized theory on a flat background
has only two tensor degrees of freedom corresponding to the
spin-two graviton of general relativity and no extra scalar degree
of freedom. Therefore, there is no consistent way to generate
primordial density perturbations in this framework due to the
lack of an extra scalar playing the role of the inflaton field.

This fact could be quite obscure since the Starobinsky theory to
which the super-renormalizable theory reduces, as well as any
$f(R)$ gravity model \cite{Sotiriou,Antonio,Clifton}, naturally
encloses an extra degree of freedom. However, this apparent
paradox is solved if one remembers that the super-renormalizable
model introduced in Ref.~\cite{M4} reduces to $R +\epsilon R^2$
gravity only after a truncation up to terms $O(1/\Lambda^4)$, where 
$\Lambda \sim 10^{-5} M_{\rm pl}$ is a parameter with energy
dimension. Meanwhile, the full theory where all orders in
$1/\Lambda^2$ are considered, which contains infinite derivatives,
does not coincide with the Starobinsky model with different
degrees of freedom. In particular the full theory has no extra
scalar degree of freedom. Therefore in some sense, the truncation
procedure adopted in Ref.~\cite{M4} is not completely consistent
with the Starobinsky theory, since it does not preserve the
degrees of freedom of the starting theory.

The aim of this paper is to show that the most general class of
super-renormalizable theories compatible with unitarity contains
at most one extra degree of freedom besides the graviton field and
reduces to the Starobincky $R+ \epsilon R^2$ theory under a
suitable truncation. The truncation is coherent in this case,
since it preserves the degrees of freedom of the theory after
truncation. Here we achieve this result and define the maximal
class of such theories; we use the term ``maximal'' to indicate that
any other extension of the theory must contain a ghost or a
tachyon and therefore unitarity is violated.

This paper is organized as follows.
In Sec.~\ref{multidimensional theory} we introduce the
super-renormalizable action for gravity in a $D$-dimensional spacetime.
In Sec.~\ref{propagator-section} we calculate the propagator for the
gravitational field fluctuation and in Sec.~\ref{unitarity and degrees of freedom}
we show that, requiring unitarity, the theory contains at most the spin-two
graviton field plus one scalar degree of freedom that we call gravi-scalar or scalaron.
In Sec.~\ref{renormalizability and finiteness} we show that the theory is
super-renormalizable in even dimension and finite in odd dimension.
In Sec.~\ref{starobisnky limit} the Starobinsky theory is recovered
through a suitable and coherent truncation of the Lagrangian density.
In Sec.~\ref{importancegravi-scalar} we expand about the importance of
the gravi-scalar degree of freedom of the theory.
Finally in Sec.~\ref{conclusions} we resume the results of
this paper and conclude.

Hereafter the spacetime metric tensor $g_{\mu \nu}$ has the signature $(+, -,
\dots, -)\,$, the curvature tensor is $R^{\mu}\!_{\nu \rho \sigma}
= - \partial_{\sigma} \Gamma^{\mu}_{\nu \rho} + \dots$, the Ricci
tensor is $R_{\mu \nu } = R^{\sigma}\!_{\mu \nu \sigma}$,  the
curvature scalar is $R = g^{\mu \nu} R_{\mu \nu}$,
Moreover we use natural units $c=1$ and $\hbar =1$.

\section{The multi-dimensional theory}\label{multidimensional theory}

In this section we introduce a general action for the class of
super-renormalizable or finite theories under consideration
in $D$-dimensional spacetime. Let us start with the following
non-polynomial or semi-polynomial Lagrangian density
\be
\hspace{-0.35cm}
\mathcal{L} &=&
2\, \kappa_D^{-2} \, R + \bar{\lambda}
+c_1^{(1)} \, R^3 +  \dots +  c_1^{(\mathrm{N})} \, R^{\mathrm{N} +2}
\nonumber \\
&&
+ \sum_{n=0}^{\mathrm{N}} \left[
a_n \, R \, (-\Box_{\Lambda})^n \, R  +
b_n \, R_{\mu \nu} \, (-\Box_{\Lambda})^n \,
R^{\mu\nu} \right]
\nonumber \\
&& + R  \, h_0( - \Box_{\Lambda}) \, R +
R_{\mu \nu} \, h_2( - \Box_{\Lambda}) \, R^{\mu\nu}
\,,\label{action}
\ee
where $\kappa_D^2 = 32 \pi G$, $G$ is the Newton's
gravitational constant, $\Box_{\Lambda} \equiv \Box/\Lambda^2$ and
the operator $\Box \equiv g^{\mu \nu} \nabla_\mu \nabla_\nu$ is
constructed with covariant derivatives. The non-polynomial
operators have been introduced in the last line of (\ref{action})
making use of the following two entire functions
\be
h_0(z) &=& - \frac{\left[ V_0(z)^{-1} -1 \right] }{\kappa_D^2 \Lambda^2 \, z}
-\sum_{n=0}^{\mathrm{N}} \tilde{a}_n \, z^n \,, \nonumber \\
h_2(z) &=& 2 \frac{\left[V_2(z)^{-1} -1 \right]}{\kappa_D^2 \Lambda^2 z}
- \sum_{n=0}^{\mathrm{N}} \tilde{b}_n \, z^n \,,
\label{hzD}
\ee
where $V_0(z)$ and $V_2(z)$ are two entire functions
that we are going to select consistently with unitarity and renormalizability.
The constants $\tilde{a}_n$ and $\tilde{b}_n$ are just non-running
parameters, while the running coupling constants are
\be
\alpha_i
\in \{\kappa_D, \bar{\lambda}, a_n, b_n, c_1^{(1)}, \cdots,
c_1^{(\mathrm{N})} \} \equiv \{\kappa_D, \bar{\lambda},
\tilde{\alpha}_n \} \label{couplings}\,.
\ee
The integer N is defined as follows in
order to avoid fractional powers of the D'Alembertian operator, namely,
\be
&& 2 \mathrm{N} + 4 = D_{\rm odd} +1 \,\,~~\mbox{in odd dimension} , \\
&& 2 \mathrm{N} + 4 = D_{\rm even} \,\,\,~~~~~~
\mbox{in even dimension}.
\ee
The ``form factors" $V_i(z)^{-1}$ $(i=0,2)$ will be
defined later for the compatibility with unitarity and renormalizability.

The goal of this paper is to find an ultraviolet completion of the Starobinsky theory
with exactly the same particle spectrum: the massless graviton and the massive gravi-scalar.
We will see later in Sec.~\ref{unitarity and degrees of freedom}
that this is the maximal particle content compatible with unitarity 
and super-renormalizability or finiteness at quantum level.
However, a non-polynomial minimal theory reproducing the Starobinsky action 
in the low-energy limit has been already introduced and studied 
in Refs.~\cite{Modesto,M4,Biswas}.
This action satisfies all the requirements 1.-5. listed in the previous section, 
but only the massless graviton propagates.
The Lagrangian in Refs.~\cite{Modesto,M4,Biswas} is the same 
as Eq.~(\ref{action}), with the identification of the two form 
factors as $V_0(z)^{-1} = V_2(z)^{-1} \equiv e^{\Box_{\Lambda}}$ \cite{Biswas, SFT}.
In Sec.~\ref{importancegravi-scalar} we see that this prescription gives rise to 
the Starobinsky Lagrangian, but the problem is that it lacks the gravi-scalar 
degree of freedom required to generate scalar metric perturbations during inflation.

\section{Propagator}
\label{propagator-section}

We shall explicitly calculate the two-point propagator
for the action (\ref{action}) and then we impose the
condition that the non-polynomial functions ${h}_0(z)$ and
${h}_2(z)$ defined in (\ref{hzD}) have to fulfill in order to
achieve a theory which satisfies the theoretical and observative
consistency requirements resumed in the introduction. We stress
that it is important here to obtain the expression of the two-point
function since from the poles of the propagator
the number of propagating degrees of freedom will be clear.

We proceed to split the spacetime metric
into the flat Minkowski background
plus a fluctuation $h_{\mu \nu}$ defined by
\be
g_{\mu \nu} =  \eta_{\mu \nu} + \kappa_D h_{\mu\nu} \, ,
\ee
and then we expand the Lagrangian to second
order in the gravitational fluctuation $h_{\mu \nu}$.
Omitting total derivative operators, we end up
with the following outcome \cite{HigherDG}
\be
&& \hspace{-0.5cm} \mathcal{L}_{\rm linear} =
- \frac{1}{2} [ h^{\mu \nu} \Box h_{\mu \nu} + A_{\nu}^2
+ (A_{\nu} - \phi_{, \nu})^2 ] \nonumber \\
&&  \hspace{-0.4cm}
+ \frac{\kappa_D^2}{8} \Big[
\Box h_{\mu \nu}  \beta( \Box) \Box h^{\mu \nu}
-A^{\mu}_{, \mu}  \beta( \Box) A^{\nu}_{, \nu}
- F^{\mu \nu}  \beta( \Box) F_{\mu \nu} \nonumber \\
&& \hspace{-0.4cm}
+ (A^{\alpha}_{, \alpha} - \Box \phi) (\beta( \Box)
+  4 \alpha( \Box) ) (A^{\beta}_{, \beta} - \Box \phi)
\Big]\,,
\label{quadratic2}
\ee
where the vector and antisymmetric tensor are below defined
in terms of the gravitational fluctuation,
\be
A^{\mu} &=& h^{\mu \nu}_{\,\,\,\, , \nu} \, , \nonumber\\
\phi &=& h^{\mu}_{\mu} \,\,~~(\mbox{trace of} \,\,h_{\mu \nu}) \, , \nonumber \\
F_{\mu \nu} &=& A_{\mu , \nu} - A_{\nu, \mu} \, ,
\ee
while the functionals of the D'Alembertian operators
$\alpha(\Box), \beta (\Box)$ are defined by
\be
&& \alpha(\Box)  :=  2  \sum_{n = 0}^{\mathrm{N} } a_n
( - \Box_{\Lambda})^n + 2 h_0(- \Box_{\Lambda}) , \nonumber \\
&& \beta(\Box)  :=  2  \sum_{n = 0}^{\mathrm{N}} b_n ( - \Box_{\Lambda})^n
+ 2 h_2(- \Box_{\Lambda})\,.
\label{alphabeta}
\ee

The d'Alembertian operator in $\mathcal{L}_{\rm linear}$ and (\ref{alphabeta})
should be evaluated on the flat spacetime.
The linearized Lagrangian (\ref{quadratic2}) is invariant under infinitesimal coordinate transformations
$x^{\mu} \rightarrow x^{\mu} + \kappa_D \, \xi^{\mu}(x)$, where $\xi^{\mu}(x)$
is an infinitesimal vector field of dimensions
$[\xi(x)] = M^{(D-4)/2}$. Under this transformation, the graviton field turns into
$h_{\mu \nu} \rightarrow h_{\mu \nu} - \xi(x)_{\mu, \nu} - \xi(x)_{\nu, \mu}.$
The presence of the local gauge symmetry
calls for the addition of a gauge-fixing term to the linearized
Lagrangian (\ref{quadratic2}). Hence, we choose the usual harmonic gauge
\be
\mathcal{L}_{\rm GF}  =
\xi^{-1} \,  A_{\mu} \,
\omega(-\Box_{\Lambda}) A^{\mu}\,,
\label{GF2}
\ee
where $\omega( - \Box_{\Lambda})$ is a gauge weight function \cite{Stelle, Tomboulis}.
The linearized gauge-fixed Lagrangian reads
\be
\mathcal{L}_{\rm linear} + \mathcal{L}_{\rm GF} =
\frac{1}{2} h^{\mu\nu} \mathcal{O}_{\mu\nu, \rho \sigma} \, h^{\rho \sigma},
\label{O}
\ee
where the operator $\mathcal{O}$ has two contributions
coming from the linearized Lagrangian (\ref{quadratic2})
and from the gauge-fixing term (\ref{GF2}).
Inverting the operator $\mathcal{O}$ \cite{HigherDG},
we find the following two-point function
\be
&& \hspace{-1cm} \mathcal{O}^{-1} = \frac{\xi (2P^{(1)} +
\bar{P}^{(0)} ) }{2 k^2 \, \omega( k^2/\Lambda^2)}
+ \frac{P^{(2)}}{k^2 \left( 1 + k^2 \kappa_D^2 \frac{\beta(k^2)}{4} \right) }
\nonumber \\
&&\hspace{-1cm}
- \frac{P^{(0)}}{k^2 \left( D-2 - k^2 \kappa_D^2\left(  D \frac{ \beta(k^2) }{4}
+ (D-1) \alpha(k^2)  \right) \right) } \, , \label{propagator}
\ee
where the projectors in $D$ dimensions are defined by \cite{HigherDG, VanNieuwenhuizen}
\be
 && \hspace{-0.2cm}
 P^{(2)}_{\mu \nu, \rho \sigma}(k) = \frac{1}{2} ( \theta_{\mu \rho} \theta_{\nu \sigma} +
 \theta_{\mu \sigma} \theta_{\nu \rho} ) - \frac{1}{D-1} \theta_{\mu \nu} \theta_{\rho \sigma} \, ,
 \nonumber
 \\
 \nonumber \\
 && \hspace{-0.2cm}
   P^{(1)}_{\mu \nu, \rho \sigma}(k) = \frac{1}{2} \left( \theta_{\mu \rho} \omega_{\nu \sigma} +
 \theta_{\mu \sigma} \omega_{\nu \rho}  +
 \theta_{\nu \rho} \omega_{\mu \sigma}  +
  \theta_{\nu \sigma} \omega_{\mu \rho}  \right) \, , \nonumber   \\
   &&
  \hspace{-0.2cm}
 P^{(0)} _{\mu\nu, \rho\sigma} (k) = \frac{1}{D-1}  \theta_{\mu \nu} \theta_{\rho \sigma}  \, , \,\,\,\, \,\,
 \bar{P}^{(0)} _{\mu\nu, \rho\sigma} (k) =  \omega_{\mu \nu} \omega_{\rho \sigma} \, ,  \,\,\,\,\,  \nonumber \\
&& \hspace{-0.2cm}
\theta_{\mu \nu} = \eta_{\mu \nu} - \frac{k_{\mu} k_{\nu}}{k^2}  \, , \,\,\,\,\,
 \omega_{\mu \nu} = \frac{k_{\mu} k_{\nu}}{k^2} \, .\label{proje2}
\ee
Note that the tensorial indices for the operator $\mathcal{O}^{-1}$
and the projectors $P^{(0)},P^{(2)},P^{(1)},\bar{P}^{(0)}$
have been omitted.\footnote{The
following identities are useful to split the terms
proportional to the gravitational momentum from the remaining:
\be
&& P^{(2)}_{\mu\nu, \rho \sigma} = \frac{1}{2} \left( \eta_{\mu \rho} \eta_{\nu \sigma} +
\eta_{\mu \sigma} \eta_{\nu \rho} \right) - \frac{1}{D-1} \eta_{\mu \nu} \eta_{\rho \sigma} \nonumber \\
&& \hspace{1cm}
- \left[ P^{(1)} + \frac{D-2}{D-1} \bar{P}^{(0)} - \frac{1}{D-1} \bar{\!\bar{P}}^{(0)}\right]_{\mu\nu, \rho \sigma} \,  , \nonumber \\
&& P^{(0)}_{\mu\nu, \rho \sigma} = \frac{1}{D-1} \eta_{\mu \nu} \eta_{\rho \sigma}
- \frac{1}{D-1} \left[\bar{P}^{(0)} + \bar{\! \bar{P}}^{(0)}  \right]_{\mu\nu, \rho \sigma}\,,
\label{identity}
\ee
where
\bee
\bar{\!\bar{P}}^{(0 )} _{\mu\nu, \rho\sigma}(k)  = \theta_{\mu \nu} \omega_{\rho \sigma}
+ \omega_{\mu \nu} \theta_{\rho \sigma} \,.
\eee
}

The functions $\alpha(k^2)$ and $\beta(k^2)$ are achieved by
replacing $-\Box \rightarrow k^2$ in the definitions
(\ref{alphabeta}).
By looking at the last two gauge-invariant terms in Eq.~(\ref{propagator}),
we deem convenient to introduce the following definitions,
\be
&& \hspace{-0.5cm}
\bar{h}_2(z) = 1 + \frac{\kappa_D^2 \Lambda^2}{2}  z \sum_{n=0}^{\mathrm{N}} b_n z^n + \frac{\kappa_D^2 \Lambda^2}{2}
z \, h_2(z) \, , \nonumber \\
&& \hspace{-0.5cm}
\bar{h}_0(z)  =  1
- \frac{\kappa_D^2 \Lambda^2 D}{2(D-2)}  z
\left[\sum_{n=0}^{\mathrm{N}} b_n z^n + h_2(z) \right] \nonumber \\
&& \hspace{0.25cm}
- \frac{2 \kappa_D^2 \Lambda^2 (D - 1)}{D-2}  z \left[\sum_{n=0}^{\mathrm{N}}a_n z^n + h_0(z) \right] , \label{barh2h0}
\ee
where again $z = - \Box_{\Lambda}$.
Through the above definitions (\ref{barh2h0}),
the propagator greatly simplifies to
\be
\hspace{-0.2cm}
\mathcal{O}^{-1}
= \frac{1}{k^2}
\left[ \frac{P^{(2)}}{\bar{h}_2}
- \frac{P^{(0)}}{(D-2) \bar{h}_0} \right]
+ \frac{\xi (2P^{(1)} +
\bar{P}^{(0)} ) }{2 k^2 \, \omega }\,.
\label{propgauge}
\ee
In the above formula we missed the argument $k^2$
for the entire functions $h_2$, $h_0$
and the weight function $\omega$.

Once established that $h_2$ and $h_0$ are not polynomial
functions, to achieve unitarity, we demand the following
general properties for the transcendental entire functions
$h_i(z)$ ($i = 0,2$) and/or $\bar{h}_i(z)$ ($i = 0,2$) \cite{Tomboulis}:
\begin{enumerate}
\renewcommand{\theenumi}{(\roman{enumi})}
\renewcommand{\labelenumi}{\theenumi}
\item $\bullet$ $\bar{h}_2(z)$ is real and positive
on the real axis and it has no zeroes on the
whole complex plane $|z| < + \infty$. \\
$\bullet$ $\bar{h}_0(z)$  is real with at most one zero
on the real axis and then at most one zero in the
whole complex plane $|z| < + \infty$.
We will show in the next section that these
requirements imply the maximal particle content compatible with unitarity.
\item $|h_i(z)|$ has the same asymptotic behavior along the real axis at $\pm \infty$.
\item There exists $\Theta>0$ such that
\be
&& \lim_{|z|\rightarrow + \infty} |h_i(z)| \rightarrow | z |^{\gamma_i
+ \mathrm{N}}, \nonumber \\
&& \gamma_i \geqslant D/2 \,\,\,\, {\rm for} \,\,\,\, D = D_{\rm even}
\,\,\,\, {\rm and} \nonumber \\
&& \gamma_i \geqslant (D-1)/2 \,\,\,\, {\rm for} \,\,\,\, D = D_{\rm odd} \, ,
\label{tombocond}
\ee
for the argument of $z$ in the following conical regions
{\small
\be
&& \hspace{0.45cm}
C = \big\{ z \, | \,\, - \Theta < {\rm arg} z < 
+ \Theta \, , \,\,  \pi - \Theta < {\rm arg} z < \pi + \Theta \big\}
\nonumber \\
&&  \hspace{0.45cm}
{\rm for } \,\,\, 0< \Theta < \pi/2. \nonumber
\ee
}
This condition is necessary to achieve the super-renormalizability
of the theory that we are going to show here below. The necessary
asymptotic behavior is imposed not only on the real axis,
but also on the conic regions that surround it.
In an Euclidean spacetime, the condition (ii) is not strictly necessary if (iii) applies.
\end{enumerate}

We regard that the theory is renormalized at some scale
$\mu_0$. Therefore, if we set
\be
\tilde{a}_n = a_n(\mu_0) \,,\qquad
\tilde{b}_n = b_n(\mu_0),
\label{betaalphaD}
\ee
we can express $\bar{h}_2(z)$ and
$\bar{h}_0(z)$ in terms of the form factors $V_2(z)$ and
$V_0(z)$ replacing (\ref{hzD}) in (\ref{barh2h0}), namely
\be
\hspace{-0.5cm}
\bar{h}_2(z)  &=& V_2(z)^{-1}
\,, \nonumber \\
\hspace{-0.5cm}
\bar{h}_0(z) &=& \frac{2}{D-2} \left[-\frac{D}{2} \, V_2(z)^{-1}
+ (D-1) V_0(z)^{-1} \right] \,.
\label{h20}
\ee
Let us assume for the moment that
the entire functions $\bar{h}_2(z)$
and $\bar{h}_0(z)$ are each a polynomial multiplied by
the exponential of an entire function, namely
\be
\bar{h}_2(z) &:=& e^{H_2(z)} p^{(n_2)}(z) \, , \nonumber  \\
\bar{h}_0(z) &:=& e^{H_0(z)} p^{(n_0)}(z) \, ,
\label{Vi}
\ee
while $p^{(n_i)}(z)$ are two polynomials of degree $n_i$
respectively. The two polynomials will be fixed
shortly in Sec. \ref{unitarity and degrees of freedom} compatibly with unitarity.
Using (\ref{h20}), we can invert (\ref{Vi}) for $V_2(z)^{-1}$ and  $V_0(z)^{-1}$,
\be
&& \hspace{-0.7cm} V_2(z)^{-1} = e^{H_2(z)} p^{(n_2)}(z) \, ,
\nonumber \\
&&  \hspace{-0.7cm}  V_0(z)^{-1} =
\frac{(D-2) e^{H_0(z)} p^{(n_0)}(z)
+ D \, V_2(z)^{-1} }{2(D-1)}\,.
\label{Vi2}
\ee

A class of entire functions $H_i(z)$ ($i=2,0$) compatible with
the required properties (i)-(iii) and the
definitions (\ref{hzD}), (\ref{Vi}) are
\be
&& \hspace{-0.4cm}
H_i(z)
= \frac{1}{2} \left[ \gamma_E +
\Gamma \left(0, p_{\gamma_i+\mathrm{N}+1}^{2}(z) \right)
+ \log \left( p^2_{\gamma_i+ \mathrm{N}+1}(z) \right) \right] \! ,
\nonumber \\
&&
\hspace{-0.4cm}
{\rm Re}\,( p_{\gamma_i+\mathrm{N}+1}^{2}(z) ) > 0 \,,
\label{HD}
\ee
where $\Gamma(a,z)$ is defined in the footnote\footnote{
$\gamma_E=0.577216$ is the Euler's constant and
\be \Gamma(a,z) = \int_z^{+ \infty} t^{a -1} e^{-t} d t
\ee is the incomplete gamma function.}
and the form factors can be written as
\be
&& \hspace{-0.5cm} e^{H_i(z)} =
e^{\frac{1}{2} \left[ \Gamma \left(0, p_{\gamma_i + \mathrm{N} +
1}^2(z) \right)+\gamma_E \right] } \, \left| p_{\gamma_i +
\mathrm{N} + 1}(z) \right|  \, .
\ee

If we choose
$p_{\gamma_i+\mathrm{N}+1}(z) = z^{\gamma_i + \mathrm{N}+ 1}$,
$H_i(z)$ simplifies to:
\be && \hspace{-0.5cm} H_i(z) = \frac{1}{2} \left[
\gamma_E + \Gamma \left(0, z^{2 \gamma_i +2 \mathrm{N}+2} \right)
+ \log (z^{2\gamma_i +2 \mathrm{N}+2}) \right] \, ,
\nonumber \\
&& \hspace{-0.5cm}
{\rm Re}(z^{2 \gamma_i +2 \mathrm{N}+2}) > 0 \,\, \,\, \Longrightarrow \,\, \,\,
\Theta = \frac{\pi}{4(\gamma_i +\mathrm{N} +1)} \, , \nonumber  \\
&& \hspace{-0.5cm}
H_i(z) = \frac{ z^{2 \gamma_i + 2 \mathrm{N}+2}}{2}
- \frac{ z^{4 \gamma_i + 4 \mathrm{N}+ 4}}{8} + \dots \,\,\, {\rm for} \,\, z \approx 0  \, ,
\label{H0}
\ee
where $\Theta$ is the angle defining
the cone $C$ of the property (iii).
The first correction to the form factor $e^{H_i(z)}$ goes to
zero faster than any polynomial function for $z \rightarrow + \infty$, namely
\be
&& \hspace{-0.5cm} \lim_{z \rightarrow +\infty} e^{H_i(z)} = e^{\gamma_E/2} \,
|z|^{\gamma_i + \mathrm{N} +1} \,,
\nonumber \\
&&\hspace{-0.5cm}
 \lim_{z \rightarrow +\infty}
\left(\frac{e^{H_i(z)}}{e^{\gamma_E/2} |z|^{\gamma_i + \mathrm{N} +1} }
- 1 \right) z^n = 0\,,\quad
\forall \, n \in \mathbb{N}\, .
\label{property}
\ee

The main result in this section is the propagator
(\ref{propgauge}) together with
the definitions (\ref{Vi}) and (\ref{HD}).

\section{Unitarity and degrees of freedom}
\label{unitarity and degrees of freedom}

In this section we discuss the unitarity of the theory (\ref{action}).
In particular we tackle the problem of defining a theory
of pure gravity with the maximal number of degrees of
freedom compatible with unitarity.

If both tachyons and ghosts are absent, the stability of the theory
is ensured at classical and quantum levels. In this case the
corresponding propagator has only first poles at
$k^2 - M_i^2 =0$ with real masses $M_i$ (no tachyons)
and with positive residues (no ghosts).

For the evaluation of the propagator (\ref{propgauge})
we make use of the explicit definitions (\ref{Vi}) of $\bar{h}_i$ ($i=2,0$)
written in terms of the entire functions $H_i(z)$ and the polynomial
functions $p^{(n_i)}$ defined through (\ref{Vi}),
\be
\hspace{-0.2cm}
p^{(n_2)}= \prod_{j=1}^{n_2} \left( 1-\frac{k^2}{\bar{m}_j^2} \right)\, , \quad
p^{(n_0)} = \prod_{j=1}^{n_0} \left( 1-\frac{k^2}{m_j^2} \right) \, ,
\label{epex}
\ee
where $\bar{m}_j^2>0$ and $m_j^2>0$.
If we couple the propagator (\ref{propgauge}) to the conserved stress-energy
tensor $T^{\mu \nu}$ satisfying the relation $\nabla_{\mu}T^{\mu \nu}=0$,
the contributions coming from the terms $P^{(1)}$ and $\bar{P}^{(0)}$
vanish from the definition (\ref{proje2}).
Dropping those contributions and using Eq.~(\ref{epex}),
the propagator (\ref{propgauge}) reads
\be
&& \hspace{-0.3cm}
\mathcal{O}^{-1}(k) =
\frac{P^{(2)}}{k^2 e^{H_2} p^{(n_2)}} -
 \frac{P^{(0)}}{(D-2) k^2  e^{H_0} p^{(n_0)}}  \label{propPoly}\\
&& \hspace{-0.3cm}
= \frac{P^{(2)} \, e^{- H_2} }{k^2  \prod_{j=1}^{n_2} \left(1- \frac{k^2}{\bar{m}_j^2 } \right)}
- \frac{P^{(0)} \, e^{- H_0}  }{(D-2) k^2  \prod_{j=1}^{n_0} \left(1- \frac{k^2}{{m}_j^2} \right)}
\nonumber \\
&& \hspace{-0.3cm}
= \frac{P^{(2)} }{ e^{H_2} }   \left(\frac{\bar{A}_0}{k^2} +
\frac{\bar{A}_1}{k^2 - \bar{m}_1^2 } + \dots +
\frac{\bar{A}_{n_2}}{k^2-\bar{m}_{n_2}^2 }
\right) \nonumber \\
&& \hspace{-0.3cm}
- \frac{P^{(0)} }{ (D-2) e^{H_0} }
\left(\frac{A_0}{k^2} +  \frac{ A_1}{k^2- m_1^2 } + \dots +
\frac{ A_{n_0}}{k^2- m_{n_0}^2 }
\right)\,, \nonumber
\ee
where $\bar{A}_j, A_j$ are constants, and
$\bar{m}_0^2=0, m_0^2=0$.

Let us assume that we have two real monotonic sequences of masses:
$\bar{m}_1< \bar{m}_2 < \dots < \bar{m}_{n_2}$,
$m_1<m_2 < \dots< m_{n_0}$.
In this case the signs of the corresponding residues alternate,
i.e., ${\rm sign}[{\rm Res}\,A_j] = -{\rm sign}[{\rm Res}\,A_{j+1}]$ and
${\rm sign}[{\rm Res}\,\bar{A}_j] = -{\rm sign}[{\rm Res}\,\bar{A}_{j+1}]$ \cite{Shapiro}.
From the propagator (\ref{propPoly}) we see that the residues
in $k^2 =0$ and $k^2 =m_1^2$ are positive but the residues
in $k^2 =\bar{m}_1^2$ and $k^2 = m_2^2$ are negative.
It follows that in order to avoid ghosts the polynomials
must have respectively degrees
\bee
n_2=0\,,\qquad n_0 \leqslant 1\,.
\eee
In fact, this meets the requirement (i) introduced and discussed
in Sec.~\ref{propagator-section}.

Let us consider the explicit example for $n_2 = 1$ and $n_0 =2$.
The spin-two and spin-zero sectors of the propagator
respectively read
\be && \hspace{-0.4cm} {\rm spin} \,\, 2 : \,
\frac{P^{(2)}}{e^{H_2} } \left( \frac{1}{k^2} - \frac{1}{k^2 -
\bar{m}_1^2} \right)\,, \\
&&  \hspace{-0.4cm}
{\rm spin} \,\, 0 : \,
\frac{P^{(0)}}{ (D-2)  e^{H_0}}
\left(-\frac{1}{k^2}+\frac{ A_1}{k^2- m_1^2 }+
\frac{ A_{2}}{k^2- m_{2}^2 } \right)\,,
\nonumber
\ee
where the constants $A_1,A_2$ are
\bee
A_1 = \frac{m_2^2}{m_2^2 - m_1^2}>0 \,,\quad
A_2 = - \frac{m_1^2}{m_2^2 - m_1^2}<0.
\eee
The quantum states have positive-definite norms and energies
if the poles in the propagator have positive residues.
In the example given above the residues in $k^2 =0$ and
$k^2 = m_1^2$ are positive, but the residues in
$k^2 = \bar{m}_1^2$ and in $k^2 = m_2^2$ are negative.

This example confirms that the maximal theory compatible with
unitarity has $n_2 =0$ and $n_0 \leqslant 1$.
The case with $n_0 =n_2 = 0$ corresponds to the model
presented in Ref.~\cite{M4}, which will be discussed with
more details in Sec.~\ref{importancegravi-scalar}.

In the case $n_2 = 0$ and $n_0 = 1$, defining $m_1^2 \equiv m^2$,
the propagator further simplifies to
\be
&& \hspace{0cm} \mathcal{O}^{-1}= \frac{P^{(2)} }{ k^2
e^{H_2} } + \frac{
m^2 P^{(0)}  }{ k^2 e^{H_0} (k^2 - m^2) (D-2) }  \\
&& \hspace{0cm} = \frac{P^{(2)} }{ k^2 e^{H_2} }   - \frac{P^{(0)}
}{ (D-2) k^2 e^{H_0} } + \frac{P^{(0)} }{ (D-2) e^{H_0} (k^2 -
m^2) } \, , \nonumber
\ee
with two single poles at $k^2=0$ and
$k^2 = m^2$ that do have positive residues because $H_i(0)=0$.
We now expand on the tree-level unitarity
coupling the propagator to external conserved stress-energy
tensors $T^{\mu \nu}$, and examining the amplitude at the poles
\cite{HigherDG, VanNieuwenhuizen}. When we introduce a general
source operator, the linearized action
is replaced by
\be
\mathcal{L}_{\rm linear} + \mathcal{L}_{\rm GF}
- g \, h_{\mu \nu} T^{\mu \nu} ,
\label{LGM}
\ee
and the transition amplitude in momentum space is
\be
\mathcal{A} = g^2 \, T^{\mu \nu} \,
\mathcal{O}^{-1}_{\mu \nu , \rho \sigma} \, T^{\rho \sigma}\,,
\label{ampli1}
\ee
where $g$ is an effective coupling constant.

To make the analysis more explicit, we can expand the sources using the
following set of independent vectors in the momentum space \cite{HigherDG},
\be
&& k^{\mu} = (k^0, \vec{k}) \, , \,\, \tilde{k}^{\mu}
= (k^0, - \vec{k}) \, , \nonumber \\
&& \epsilon^{\mu}_i = (0, \vec{\epsilon}_i) \, ,
\,\, i =1, \dots , D-2 \, ,
\ee
where $\vec{\epsilon}_i$ are $D-2$ unit vectors
orthogonal to each other and to $\vec{k}$,
\be
\vec{k} \cdot \vec{\epsilon}_i = 0 \,\, \Rightarrow \,\,
k_{\mu} \epsilon^{\mu}_i = 0 \, , \,\,\,
\vec{\epsilon}_i \cdot \vec{\epsilon}_j = \delta_{i j} .
\ee
The most general symmetric stress-energy tensor can be
expressed as
\be
&& T^{\mu\nu} = a k^{\mu} k^{\nu} + b \tilde{k}^{\mu} \tilde{k}^{\nu}
+ c^{i j} \epsilon_i^{(\mu} \epsilon_j^{\nu)} + d \, k^{(\mu} \tilde{k}^{\nu)}
\nonumber \\
&& \hspace{1cm}
+ e^i k^{(\mu} \epsilon_i^{\nu)} + f^i \tilde{k}^{(\mu} \epsilon_i^{\nu)}\,,
\ee
where we used the notation $a_{(\mu} b_{\nu)} \equiv (a_{\mu} b_{\nu}+b_{\mu} a_{\nu})/2$.
The conditions $k_{\mu} T^{\mu \nu} =0$ and $k_{\mu} k_{\nu}T^{\mu \nu} =0$ provide
the following constraints and consistency conditions
on the coefficients $a,b,d, e^i, f^i$ \cite{HigherDG}:
\be
&& \hspace{-0.5cm}
\bullet \,\,\, k_{\mu} T^{\mu \nu} =0 \,  \Longrightarrow \,
 \left\{ \begin{array}{lll}
         a k^2 + d(k_0^2 + \vec{k}^2)/2 =0\,,  \\
        b (k_0^2 + \vec{k}^2) + dk^2/2 =0\,, \\
        e^i k^2 + f^i (k_0^2 + \vec{k}^2) = 0\,.
        \end{array} \right\}
        \label{C1} \\
&&   \hspace{-0.5cm}  \Longrightarrow \, \,
        d=0 \,  , \,\, b = 0 \, , \,\,  f^i = 0  \,\,\,
\mbox{for $k^2:=k_0^2-\vec{k}^2=0$} \, ,
\label{C2} \\
&& \hspace{-0.5cm}
\bullet \,\,\,  k_{\mu} k_{\nu} T^{\mu \nu} =0  \,\,
(\mbox{consistency relation for $a$, $b$, and $d$})
\nonumber  \\
&& \hspace{-0.5cm}
\Longrightarrow \,\,
a k^4 +b (k_0^2 + \vec{k}^2)^2 + d  k^2 (k_0^2 + \vec{k}^2) = 0\,.
\label{C3}
\ee

Introducing the spin-projectors and making use of the identities
(\ref{identity}) together with the conservation of the stress-energy tensor
$k_{\mu} T^{\mu \nu} = 0$, the amplitude (\ref{ampli1}) for $n_2 =0$
and $n_0 =1$ reads
\be
&& \hspace{-0.8cm}
A = g^2 T^{\mu \nu} \left[ \frac{P^{(2)}_{\mu \nu, \rho \sigma}}{k^2 e^{H_2} p^{(n_2)}}
- \frac{P^{(0)}_{\mu \nu, \rho \sigma}}{(D-2) k^2  e^{H_0} p^{(n_0)}} \right]
T^{\rho \sigma} \nonumber \\
&& \hspace{-0.8cm}
= g^2  \left[  \frac{T_{\mu \nu} T^{\mu\nu} - \frac{T^2}{D-1}} {k^2 e^{H_2} }
- \frac{\frac{T^2}{D-1}}{ (D-2) k^2  e^{H_0} \left( 1 - \frac{k^2}{m^2} \right)  } \right] \! ,
\label{ampli2}
\ee
where $T:=\eta^{\mu \nu}T_{\mu \nu}$.

We now calculate the residue of the amplitude in $k^2 =0$ and $k^2 = m^2$.
Using the properties $H_i(0) = 0$ and (\ref{C2}), the residue
in $k^2=0$ reads
\be
&& \hspace{-0.5cm}
{\rm Res}\,A\Big|_{k^2 = 0} = g^2 \! \left(  T_{\mu \nu} T^{\mu \nu}
- \frac{T^2}{D-2} \right) \Bigg|_{k^2 = 0}
\nonumber \\
&& \hspace{-0.5cm}
= g^2 \! \left[ (c^{ij})^2 - \frac{ (c^{ii})^2}{D-2} \right] \Bigg|_{k^2 =0}  \!\!\!\!\!\! > 0
\,\,\,\,  {\rm for} \,\, \,\, D>3.
\label{residuo}
\ee
When $D=3$, the graviton is not a dynamical degree of freedom
and the amplitude is zero.
The residue in $k^2 = m^2$ results
\be
\hspace{-0.4cm}
 {\rm Res}\,A\Big|_{k^2 = m^2} =
 g^2 \frac{T^2 e^{- H_0(m^2/\Lambda^2)}}{(D-1)(D-2)} 
> 0 \,
\,\, \, {\rm for} \,\,\,  D>2\,,
\label{residuo2}
\ee
in which case the scalar mode propagates. Thus, in the case $n_2 =
0$ and $n_0 = 1$, the spectrum consists of two particles: the
graviton and the gravi-scalar (scalaron).  We conclude that the
maximal class of super-renormalizable unitary theories
includes a gravi-scalar besides the graviton.

\section{Renormalizability and Finiteness}
\label{renormalizability and finiteness}

In this section we study the power counting renormalizability of
the theory, showing that it is renormalizable in even
spacetime dimension and finite in odd dimension.

The theory can be renormalizable if we assume the same ultraviolet
behavior for the function $\bar{h}_2(z)$ and $\bar{h}_0(z)$.
For $n_2 =0$ and $n_0=1$, it follows that
\be
&& \bar{h}_2(z) = e^{H_2} \,\, \rightarrow  \,\, z^{\gamma_2 + \mathrm{N} +1}\,, \,\,\,\,
{\rm and} \nonumber \\
&& \bar{h}_0(z) = e^{H_0} \left( 1 - \frac{\Lambda^2 z}{m^2 }\right) \,\, \rightarrow \,\,
z^{\gamma_0 + \mathrm{N} +2} \,.
\ee
If $\gamma_2 = \gamma_0 +1 \equiv \gamma$, then
the functions $\bar{h}_2(z)$ and  $\bar{h}_0(z)$
have the same scaling property.

Replacing $\gamma_2 =\gamma$ and $\gamma_0 = \gamma -1$
and using Eqs.~(\ref{action}), (\ref{hzD}), (\ref{barh2h0}),
and (\ref{propgauge}),
the high-energy scaling of the propagator in the momentum space
and the leading interaction vertex are schematically
given by
\be
&& \hspace{-0.5cm}
\mathcal{O}^{-1}(k) \sim \frac{1}{k^{2 \gamma +2 \mathrm{N} +4}} \,\,\,\,\,\,
\mbox{in the ultraviolet} \,,
\label{intera3} \\
&& \hspace{-0.5cm}
{\mathcal L}^{(n)} \sim  h^n \, \Box_{\eta} h \,\,  h_i( - \Box_{\Lambda}) \,\, \Box_{\eta} h
\,\,\, \rightarrow \,\,\,   h^n \, \Box_{\eta} h
\,   \Box_{\eta}
^{\gamma + \mathrm{N} } \,
\Box_{\eta} h\,,
\nonumber
\ee
where $\Box_{\eta} := \eta^{\mu\nu} \partial_{\mu} \partial_{\nu}$.
In (\ref{intera3}) the indices for the gravitational fluctuations
$h_{\mu \nu}$ are omitted (replaced by $h$),
and $h_i( - \Box_{\Lambda})$ is the entire function
defined by the properties (i)-(iii).
From (\ref{intera3}),
the upper bound to the superficial degree of divergence in a
spacetime of ``even" or ``odd" dimension is
\be
\hspace{-0.4cm}
w(G)_{\rm even} &=& D_{\rm even} - 2 \gamma  (L - 1) \,, \nonumber \\
\hspace{-0.4cm} 
w(G)_{\rm odd} &=& D_{\rm odd} - (2 \gamma+1)  (L - 1)\,,
\label{diverE}
\ee
where we used the topological relation between vertexes $V$, internal lines $I$ and
number of loops $L$: $I = V + L -1$.
Thus, if $\gamma > D_{\rm even}/2$ or $\gamma > (D_{\rm odd}-1)/2$,
only 1-loop divergences survive in this theory.
Therefore, the theory is super-renormalizable, unitary and
microcausal as pointed out also in Refs.~\cite{Krasnikov, efimov, E3, E4, E5}.
For $\gamma$ sufficiently large the divergent contributions to
the $\beta$-functions ($\beta_i$) are independent
from the running coupling constants (\ref{couplings}) and then
the $\beta$-functions do not depend on the energy scale $\mu$
defined trough $t := \log \left(\mu/\mu_0 \right)$.
It follows that we can easily
integrate the renormalization
group equations \cite{Modesto}, i.e.,
\be
\frac{d \alpha_i}{d t} = \beta_i  \,\,\,\, \Longrightarrow \,\, \,\,
\alpha_i(t) \sim \alpha_i(t_0) + \beta_i  t \, .
\label{also}
\ee

The mass of the gravi-scalar is not subject to renormalization and the logarithmic quantum corrections
to the propagator leave its value almost invariant because the damping factor $e^{-H_0(z)}$
suppresses any high energy shift, namely
\be
&& \hspace{-0.4cm} \mathcal{O}^{-1} =
\frac{P^{(2)} e^{-H_2}}{ k^2
 \left[1+ e^{- H_2 }    k^2 \left(c_0  + \dots + c_{\rm N} k^{2 \mathrm{N} } \,
\right) \log \left( \frac{k^2}{\mu^2} \right)
 \right] }
  \nonumber \\
  && \hspace{-0.3cm}
  + \frac{
m^2 P^{(0)} e^{- H_0 }  }{ (D-2) k^2
\left[(k^2 - m^2)+ e^{- H_0 }
k^2 \left(\bar{c}_0  + \dots
\right) \log \left( \frac{k^2}{\mu^2} \right)
 \right]},
\nonumber
\ee
where $c_0, \bar{c}_0\dots, c_{\mathrm N}$ are
dimensionfull constants and $\mu$ is a renormalization group
invariant sale.

However, in {\em odd} dimension there are no local invariants (using dimensional regularization)
with an odd number of derivatives which could serve as counter-terms for pure gravity.
This is a consequence of the rational nature of the entire functions which characterize
the theory (one example of non rational function is $h_i(\sqrt{z})$).
We conclude that all the amplitudes with an arbitrary number of loops are finite
and all the beta functions are identically zero in odd dimension,
\be
&& \beta_{a_n} = \beta_{b_n} = \beta_{c_i^{(n)}} = 0\,,  \nonumber \\
&& i \in \{1, \dots, ({\rm number \,\, of \,\, invariants \,\, of \,\, order} \,\, \mathrm{N})\} \, , \nonumber \\
&&  n =1, \dots, \mathrm{N} .
\label{betaf}
\ee
It follows that we can fix to zero all the couplings $c_i^{(n)}$
and set to constants the couplings $a_n(\mu)$ and
$b_n(\mu)$, namely
\be
&& c_i^{(n)} = 0 \,  ,\nonumber \\
&& a_n(\mu) = {\rm constant}= \tilde{a}_n \,,
\nonumber \\
&& b_n(\mu) = {\rm constant}= \tilde{b}_n  \,.
\label{costanti}
\ee

Therefore, quantum gravity is finite in even dimension, as
well, once the Kaluza-Klein compactification is applied
\cite{duff}. The finiteness of the theory in even dimensions
follows from the inclusion of an infinity tower of states which
drastically affects the ultraviolet behavior.
%
%Applying the results in \cite{Shapiro} to the theory (\ref{action}) with $n_0 =1$ and $n_2 =0$
%it is easy to show that the beta function for the cosmological constant can be fixed to zero
%for particular values of the non-running free parameters in the theory. It follows that we can set the
%value of the cosmological constant consistently with the observative value.

%
\section{Starobinsky limit}\label{starobisnky limit}

In the following we show how the Lagrangian (\ref{action})
reduces to the Starobinsky $R + \epsilon R^2$ theory after a
suitable truncation for large values of the $\Lambda$ parameter.
We also discuss how to fix the value of such a mass scale and
the value of the gravi-scalar mass $m$.

The Lagrangian (\ref{action}) can be recast as follows
\be \!\!\! && \hspace{-0.3cm} \mathcal{L} = \frac{2}{\kappa_D^2}
\left( R - G_{\mu \nu} \frac{V^{-1}_2 - 1}{\Box}
 R^{\mu \nu}  + \frac{1}{2}R \frac{V^{-1}_0 - V^{-1}_2}{ \Box} R \right)  ,
\nonumber \\
&& \hspace{-0.3cm} V^{-1}_0 - V^{-1}_2 = \frac{D-2}{2(D-1)} \left[
e^{H_0} \left( 1 + \frac{\Box}{m^2} \right) - e^{H_2} \right]\,,
\label{actionF}
\ee
where $G_{\mu \nu}$ is the Einstein's tensor.
Expanding the above Lagrangian (\ref{actionF})
for large $\Lambda$, we find
\be
\!\!\!\! \mathcal{L} = \frac{2}{ \kappa_D^{2}} \left[ R +
\frac{(D-2)R^2 }{4 (D-1) m^2} + O \! \left( R \frac{\Box^{2
\gamma + 2 \mathrm{N} -1}}{\Lambda^{4 \gamma + 4 \mathrm{N}}} R \!
\right) \right] \! . \label{StaroLimit}
\ee
When
\be
R \frac{\Box^{2 \gamma + 2 \mathrm{N} -1}}
{\Lambda^{4 \gamma+ 4 \mathrm{N}}}R \ll
\frac{(D-2)R^2 }{4 (D-1) m^2}\,,
\label{stalimitcon}
\ee
the last term in (\ref{StaroLimit}) is negligible and in $D=4$ dimensions the
Lagrangian (\ref{StaroLimit}) reduces exactly to (\ref{Staro})
with $\epsilon =1/(6m^2)$.

It seems natural to identify $\Lambda$ and the gravi-scalar mass
$m$ to avoid a further mass scale in the classical theory.
Unlike the previous model \cite{M3}, we obtain the Starobinsky
$R+\epsilon R^2$ theory with exactly the same spectrum, the massless graviton
and the massive gravi-scalar essential to generate proper
primordial density perturbations.

The equation of motion up to operators $O(R\Box R)$
and $O(R^3)$ reads \cite{barvy, barvy2, barvy3},
\be
&&
\hspace{-1cm}
G_{\mu \nu} + \frac{D-2}{2(D-1)m^2} R\left( R_{\mu \nu} - \frac{1}{4} g_{\mu \nu} R\right)
\nonumber \\
&& \hspace{-1cm} -  \frac{D-2}{2(D-1)m^2} \left( g_{\mu \nu} \Box
- \nabla_{\mu} \nabla_{\nu} \right) R = \frac{\kappa_D}{4}
T_{\mu\nu} \,, 
\ee 
which is exactly the Starobinsky equation of
motion in $D$ dimensions\footnote{
In calculating the variation of the action
(\ref{actionF}) we used the compatibility property of the metric
$\nabla_{\mu} g_{\rho \sigma} =0$ and the following variation of
the Ricci tensor, \be && \hspace{-0.5cm}
\delta R_{\mu \nu} = - \frac{1}{2} g_{\mu \alpha} g_{\nu \beta} \Box \delta g^{\alpha \beta} + \nonumber \\
&& \hspace{-0.5cm} - \frac{1}{2} \big[ \nabla^{\beta} \nabla_{\mu}
\delta g_{\beta \nu} + \nabla^{\beta} \nabla_{\nu} \delta g_{\beta
\mu} - \nabla_{\mu} \nabla_{\nu} \delta g_{\alpha \alpha} \big] \, ,
\ee
together with $\nabla^{\mu} G_{\mu \nu} =0$. The Starobinsky
action is manifestly generally covariant. Therefore, its
variational derivative exactly satisfies the Bianchi identity.}.

\begin{figure}
\hspace{-0.3cm}
\includegraphics[height=3.2in,width=3.3in]{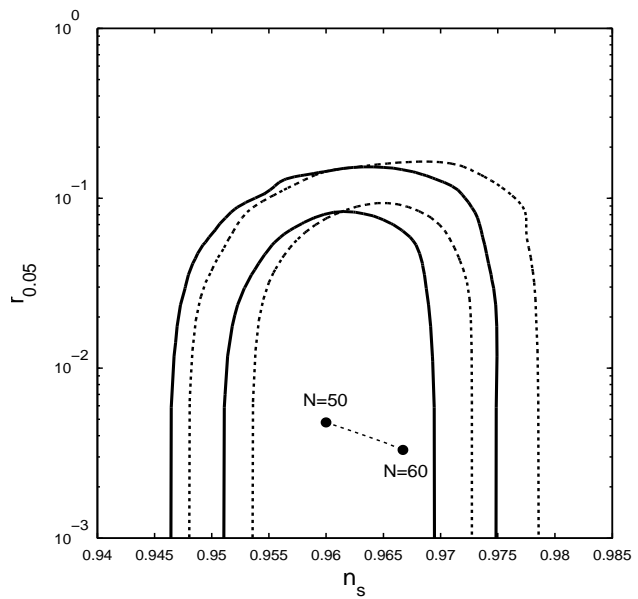}
\caption{\label{fig1}
2-dimensional observational constraints on the Starobinsky's
inflationary model ${\cal L}=(2/\kappa_D^2)[R+R^2/(6m^2)]$
in the $(n_s,r)$ plane
with the pivot wave number $k_0=0.05$~Mpc$^{-1}$.
The bold solid curves show the 68\,\%\,CL (inside) and
95\,\%\,CL (outside) boundaries derived by
the joint data analysis of Planck+WP+BAO+high-$\ell$,
while the dotted curves correspond to the 68\,\% and
95\,\% contours constrained by Planck+WP+BAO.
We plot the theoretical predictions for $N=50, 60$
as black points.
}
\end{figure}

Let us discuss observational signatures for the model
described by the Lagrangian (\ref{StaroLimit}) with $D=4$ under
the condition (\ref{stalimitcon}), i.e.,
\be
{\cal L}=\frac{2}{\kappa_D^2} \left( R+\frac{R^2}{6m^2} \right).
\ee
The density perturbations generated in the inflationary models
based on $f(R)$ gravity and scalar-tensor theory were
studied in detail in Refs.~\cite{fRper}.
The resulting power spectra of curvature perturbations ${\cal R}$
and gravitational waves $h_{ij}$ are given, respectively,
by \cite{Antonio}
\bee
{\cal P}_{\cal R}=\frac{N^2}{24\pi^2} \left( \frac{m}{M_{\rm pl}}
\right)^2\,,\qquad
{\cal P}_{h}=
\frac{1}{2\pi^2} \left( \frac{m}{M_{\rm pl}}
\right)^2\,,
\eee
where $N$ is the number of e-foldings from the end of
inflation to the epoch at which the perturbations
relevant to the CMB anisotropies with the physical wave length $a/k$
($a$ is the scale factor) crossed the Hubble radius $H^{-1}$.
From the Planck normalization ${\cal P}_{\cal R}=2.2 \times 10^{-9}$
at the pivot comoving wave number $k_0=0.05$ Mpc$^{-1}$ \cite{Tsujikawa},
the scalaron mass is constrained to be
\bee
\frac{m}{M_{\rm pl}}= 1.3 \times 10^{-5}
\left(\frac{55}{N} \right)\,,
\label{Lambdaconstrain}
\eee
which corresponds to $m \simeq 3.2 \times 10^{13}$~GeV
for $N=55$.
Since the scalaron is very heavy, the fifth force is strongly suppressed
in the present Universe. Hence the model is compatible with
local gravity constraints in the solar system \cite{Antonio,Capo}.

The scalar spectral index
$n_s-1 \equiv d\ln {\cal P}_{\cal R}/d\ln k|_{k=aH}$ reads \cite{Antonio}
\bee
n_s -1=-\frac{2}{N}=-3.6 \times 10^{-2} \left(\frac{55}{N} \right)\,,
\eee
whereas the tensor-to-scalar ratio $r \equiv {\cal P}_h/{\cal P}_{\cal R}$ is
\bee
r =\frac{12}{N^2}=4.0 \times 10^{-3} \left(\frac{55}{N} \right)^2 \,.
\eee
In order to test the observational viability of the model,
we run the CosmoMC code \cite{cosmomc,Lewis}
by setting the runnings of the scalar and tensor spectral indices to be 0.
The flat $\Lambda$CDM model is assumed with $N_{\rm eff} = 3.046$
relativistic degrees of freedom and with the instant reionization.
In Fig.~\ref{fig1} we plot the 68\,\%\,CL and 95\,\%\,CL boundaries
(solid curves) constrained by the joint analysis of Planck \cite{Planck1},
WP \cite{WMAP9}, Baryon Acoustic Oscillations (BAO) \cite{BAO}, and high-$\ell$
ACT/SPT temperature data \cite{Das} (solid curves),
together with the boundaries constrained
by Planck+WP+BAO (dotted curves).
We also show the theoretical values of $n_s$ and $r$
for $N$ between 50 and 60.
The model is well inside the 68\,\%\,CL
observational boundaries.

We also note that, in the Starobinsky's model, the non-linear parameter
$f_{\rm NL}$ of scalar non-Gaussianities is much smaller
than $1$ \cite{fRnonGau}.
This is consistent with the recent bounds of $f_{\rm NL}$ constrained
by the Planck group \cite{Planck4}.

In the next section we will argue more on the importance and the
physical implications of the gravi-scalar degree of freedom.

\section{Importance of the gravi-scalar degree of
freedom}\label{importancegravi-scalar}

Let us discuss the reason why it is important to
consider super-renormalizable and unitary theories with one
extra gravitational scalar degree of freedom.

The Starobinsky model, as well as any $f(R)$ theory
\cite{Sotiriou,Antonio,Clifton}, contains an extra scalar degree
of freedom responsible for the generation of primordial density
perturbations. This is evident after mapping the theory from
the Jordan to the Einstein reference frame in which the scalaron
has a nearly flat potential to drive inflation \cite{Antonio}.
Therefore, such a scalar plays a fundamental role for the
construction of a coherent cosmological model and,
if it is absent, one has to resort to different mechanisms to
generate primordial perturbations, e.g., Higgs inflation with
non-minimal couplings \cite{Bezrukov}.

In Ref.~\cite{M4} a super-renormalizable model
characterized by the following Lagrangian density
has been proposed
\bee
\mathcal{L} =  R - G^{\mu \nu}
\left(\frac{V(\Box_{\Lambda})^{-1} -1}{\Box} \right)R_{\mu\nu} \,,
\label{compact}
\label{FFS}
\eee
with the specific choice
\bee
V(\Box_{\Lambda}) \equiv \exp(-\Box_{\Lambda}).
\eee
This corresponds to the Lagrangian (\ref{action}) with the choice
$V_0(\Box_{\Lambda}) = V_2(\Box_{\Lambda}) = V(\Box_{\Lambda})$,
see Eq.~(\ref{actionF}).
Then, the propagator of the gravitational field
on a Minkowskian background reads
\be
\mathcal{O}^{-1} =
\frac{V(k^2/\Lambda^2)  } {k^2}
\left( P^{(2)} - \frac{P^{(0)}}{D-2} \right)\,.
\label{propgauge2}
\ee
Hence one has only the spin-two massless
graviton and no gravi-scalar degree of freedom.
The Lagrangian in this model is given by
\bee
\mathcal{L} = \frac{2}{\kappa_D^2}
\left[ R + \frac{R^2}{6\Lambda^2}
+ O\left( \frac{R \, \Box R}{ \Lambda^4} \right)
\right]\,,
\label{TRU}
\eee
which reduces to 
\be
\mathcal{L}=
2 \kappa_D^{-2}\left( R+\frac{R^2}{6\Lambda^2} \right) \, , 
\ee
for $R  \Box R/\Lambda^2 \ll R^2$.
However this reduction is not coherent with inflation since,
as we have shown above, the theory (\ref{compact}) contains
only the spin-two graviton while the Starobinsky model includes
an additional scalar degree of freedom,
so the full theory starting from the Lagrangian (\ref{compact})
and the ``reduced" one have different degrees of freedom.
This means that the truncation of the $O(R \Box R/\Lambda^4)$
terms from the Lagrangian (\ref{TRU}) is not consistent.

In Sec.~\ref{unitarity and degrees of freedom} we constructed
super-renormalizable and unitary theories containing one scalar
degree of freedom, by which gravitation is responsible for both the
inflationary expansion and the generation of perturbations.
These theories, after the
truncation of (\ref{actionF}), reduce in a coherent way to the
Starobinsky model, since the number of degrees of freedom is
preserved in the truncation procedure. Therefore, in such a case,
the gravi-scalar generates primordial perturbations and its mass
has to fulfill the condition (\ref{Lambdaconstrain}) in order to
give the correct amplitude of primordial perturbations. In
Sec.~\ref{unitarity and degrees of freedom} we have shown that the
maximal class of unitary theories verifying this requirement
contains only one extra scalar, since if any other degree of
freedom is present, it must be a ghost or a tachyon.
Hence the theory presented here is the maximal one.

\section{String Field theory}\label{string theory}

In this section we show how the Starobinsky theory emerges from
string theory when the modifications suggested by ``string field theory"
are taken into account.

In string field theory the propagator of the point-particle effective 
field theory is modified to \cite{SFT}
\be
\frac{1}{\Box} \rightarrow \frac{e^{- \tilde{\alpha}^{\prime}  \,
\Box}}{\Box}\,,
\ee
where $\tilde{\alpha}^{\prime}\equiv (\alpha^\prime/2)\ln(3\sqrt{3}/4)
\approx 0.1308\alpha^\prime$ with $\alpha^\prime$ being the universal Regge
slope parameter of the string.
Collecting together the modification suggested by string field theory and general
covariance, we propose the following effective Lagrangian
for the bosonic sector of string theory,
\be
& &\mathcal{L}_{\rm string-field} = 2 \kappa^{-2}_D  \left(R- G_{\mu\nu}\,
\frac{e^{ \tilde{\alpha}^{\prime}  \,
\Box} -1}{\Box}\, R^{\mu\nu}\right)
\nonumber \\
&&
+ \frac{1}{2}  \nabla_\mu \phi \, e^{ \tilde{\alpha}^{\prime}  \, \Box} \, \nabla^\mu \phi
+ \frac{1}{2 n !} e^{c \phi} \, F_{[n]} \, e^{\tilde{\alpha}^{\prime}  \, \Box} \, F_{[n]}  \,.
\label{stringA}
\ee
This is confirmed by the analysis in Ref.~\cite{GrossWitten},
where the authors make a field redefinition compatible with our proposal.
Let us now consider the low-energy expansion
of the exponential form factor in the Lagrangian (\ref{stringA}),
that is
\be
e^{ \tilde{\alpha}^{\prime}  \, \Box} \approx 1+
 \tilde{\alpha}^{\prime} \Box + O((\tilde{\alpha}^{\prime}\Box)^2).
\ee
The gravity sector of the Lagrangian (\ref{stringA})
simplifies to
\be
\hspace{-0.5cm}
\mathcal{L}_{\rm string-field} &\simeq&
2 \kappa^{-2}_D  \left(R- \tilde{\alpha}^{\prime} G_{\mu\nu}R^{\mu\nu}\right) \nonumber \\
\hspace{-0.5cm}
&=& 2 \kappa^{-2}_D  \left(R- \tilde{\alpha}^{\prime} R_{\mu\nu}R^{\mu\nu} + \frac{1}{2} \tilde{\alpha}^{\prime} R^2 \right)\,.
\label{stringAL}
\ee
In $D=4$, for the Friedmann-Lema\^{i}tre-Robertson-Walker metric,
the following term turns out to be topological,
\be
\int d^4 x \sqrt{|g|} \left(3 R_{\mu
\nu} R^{\mu \nu} - R^2 \right) = {\rm topological} \,,
\ee
so that the truncated theory (\ref{stringAL}) reads
\be
\hspace{-0.2cm}
\mathcal{L}_{\rm string-field} =  2 \kappa^{-2}_D
\left[ R + \frac{\tilde{\alpha}^{\prime} }{6} R^2
+ O\left( ( \tilde{\alpha}^{\prime})^2 R \, \Box R )
\right) \right] \! .
\label{TRU2}
\ee
When $\tilde{\alpha}^{\prime} R \Box R \ll R^2$, the above
Lagrangian reduces to the Starobinsky  model (\ref{Staro}).

Finally, ``string field theory" offers an alternative completion
of the Starobinsky inflation. In order to obtain all the results
of the perturbation spectra compatible with the Planck data, we need
to identify and select out an extra scalar degree of freedom from the string spectrum.
However, this is not an easy task as for the theory (\ref{actionF})
where the scalar field is a part of the gravitational sector.

In this model the gravi-scalar does not appear unlike
the study in Sec.~\ref{unitarity and degrees of freedom}
and the situation is exactly the same as the one
discussed in Sec.~\ref{importancegravi-scalar}.
The difficult task in string theory is to pull out the extra scalar degree of freedom
from the string spectrum.

%\vspace{0.5cm}

%
\section{Conclusions}
\label{conclusions}

In this paper we proposed and extensively studied
a class of super-renormalizable or finite theories of gravity
which provide an ultraviolet completion of the Starobinsky theory.
This class of theory is a generalization of the study done previously in searching for unitary
and perturbatively consistent theory of quantum gravity \cite{Modesto, Krasnikov, Tomboulis}.
The outcome is universal once few observative and theoretical assumptions have been made.

If we require the hypothesis 1.-5. listed in the introduction,
together with the validity of perturbative theory, then we find only two unitary
and super-renormalizable or finite theories of gravity.
The minimal one contains only the graviton, but it is shown that
a maximal extension is viable containing one extra scalar
degree of freedom (gravi-scalar or scalaron).
This is fundamental to generate primordial density perturbation
during inflation. The result is two-fold, on the one
hand in this theory the graviton and gravi-scalar fill up the
maximal particle content compatible with unitarity and
renormalizablity or finiteness, on the other hand the Starobinsky
model is coherently achieved in the low-energy limit.

We also mentioned other theories capable to give a completion
of the Starobinsky model.
We expound about ``asymptotic safe quantum gravity"
where at non-perturbative level the ghost pole is moved to infinity
by the renormalization group. In the string theory framework,
we studied a point-particle theory incorporating the ``string field theory"
modified propagator together with general covariance.
The resulting effective theory is in our class of super-renormalizable
theories for a particular choice of the form factor and reduces to the
Starobinsky model at low energy.

Finally, we believe the effort made in the search for a completion of
quadratic gravity to be relevant and pertinent in the light of
the recent Planck data supporting the Starobinsky
inflation \cite{Planck4,Tsujikawa}.
We would like to invite expert readers to invest
time in this research. Our instinct is confirmed by recent papers
having the same aim of
this \cite{Cecotti1987,Cecotti1988,Ketov,Watanabe,Ellis,Kallosh,sugra1,sugra2,sugra3,Ferrara2013}.

\vspace{-0.6cm}
\begin{acknowledgments}
L.~M. and S.~T. acknowledge the i-Link cooperation program
of CSIC (project ID i-Link0484) for partial sponsorship.
F.~B. is a Marie Curie fellow of the Istituto
Nazionale di Alta Matematica Francesco Severi.
S.~T. is supported by the Scientific Research Fund of the
JSPS (No.~24540286) and financial support from Scientific Research
on Innovative Areas (No.~21111006).
We thank Junko Ohashi for the help to run the CosmoMC code.
\end{acknowledgments}

\end{document}